# Enhancing community detection by local structural information

Ju Xiang[1,2,3], Ke Hu[4,*], Yan Zhang[5], Mei-Hua Bao[1,2,3], Liang Tang[1,2,3], Yan-Ni Tang[3], Yuan-Yuan Gao[3], Jian-Ming Li[1,2,*], Benyan Chen[4] and Jing-Bo Hu[4]

[1]Neuroscience Research Center, Changsha Medical University, Changsha, 410219, Hunan, China
[2]Department of Anatomy, Histology and Embryology, Changsha Medical University, Changsha, 410219, Hunan, China
[3]Department of Basic Medical Sciences, Changsha Medical University, Changsha, 410219, Hunan, China
[4]Department of Physics, Xiangtan University, Xiangtan 411105, Hunan, China
[5]Department of Computer Science, Changsha Medical University, Changsha, 410219, Hunan, China

**Abstract**:
Many real-world networks such as the gene networks, protein-protein interaction networks and metabolic networks exhibit community structures, meaning the existence of groups of densely connected vertices in the networks. Many local similarity measures in the networks are closely related to the concept of the community structures, and may have positive effect on community detection in the networks. Here, various local similarity measures are used to extract the local structural information and then are applied to community detection in the networks by using the edge-reweighting strategy. The effect of the local similarity measures on community detection is carefully investigated and compared in various networks. The experimental results show that the local similarity measures are crucial to the improvement for the community detection methods, while the positive effect of the local similarity measures is closely related to the networks under study and the applied community detection methods.

# Contents

* Corresponding author: Ke Hu and Jian-Ming Li.
Email: huke1998@aliyun.com (Ke Hu); ljming0901@sina.com (Jian-Ming Li); xiang.ju@foxmail.com, xiangju@aliyun.com (Ju Xiang).

## 1. Introduction

Community detection is a common challenge in the study of complex networks [1]. Many real-world networks such as the gene networks, protein-protein interaction networks and metabolic networks exhibit community structures—groups of vertices that are densely interconnected while only sparsely connected with the rest of the network. It is widely believed that the ability to detect such communities in networks could be is of considerable interest for understanding the structures and functions of the complex networks, because they generally correspond to such functional units as the functional modules in the protein-protein interaction networks and the cycles and pathways in the metabolic networks [1-3].

Many community detection methods have been proposed, which are based on various approaches such as spectral analysis [4, 5], random walk dynamics [6, 7], (de-)synchronization [8-10], label propagation [11-14], statistical models [15, 16], and others [17-19] (see Refs [1, 20, 21] for reviews). Most of methods depend on the quality functions for community detection, such as *modularity* [22]. Modularity optimization has been a kind of popular strategy for analyzing the community structures in networks [23-28]. While it was found to encounter some problems, especially the resolution limit, that is, small communities below a certain scale will be undetectable [29-31]. To attack the resolution limit problem, many methods have been proposed, such as the recursive partitioning of networks [32], and the edge-reweighting strategy [33].

In ref [34], the authors showed that, by using the local structural information to reconstruct the (structural) weight of existing edges in networks, the existing community detection method could identify the communities more effectively in the weighted networks than in the original networks. The edge-reweighting strategy is very simple and effective, and could directly apply to the existing community detection methods. In recent years, the edge-reweighting strategy again exhibits its powerful witchery in enhancing community detection [19, 33, 35]. Indeed, the appropriate edge-reweighting strategy could bring about the improvement of community detection, if the edge-reweighting strategy could increase the weight of the intra-community edges while decrease the weight of the inter-community edges, because: (a) this will increase the identifiability of the inter- and intra-community edges, making the community structures in networks more obvious, accelerating the effective search of the community structures in networks; (b) this leads to the relative decrease of the weight between communities, mitigating therefore the resolution limit problem of modularity. So the edge-reweighting is an effective strategy for enhancing the ability of the community detection methods in networks.

In general, the groups of vertices with dense links and strong interactions correspond to the groups of vertices with similar properties or functions, vice versa. In this sense, many local similarity measures in networks are closely related to the concept of the community structures. They may have positive effect on community detection in networks. Here, a variety of local similarity measures, by using the edge-reweighting strategy, will be applied to community detection in networks. Concretely, we will use various local similarity measures to extract the local structural information in networks, and then preprocess the networks, this is to say, reweight the edges in networks by the local structural information. In this paper, we will focus on how these local similarity measures affect the community detection in different networks, and systematically compare the effect of different local similarity measures in the networks.

## 2. Method

### 2.1 Brief review of local similarity measures

The communities are often defined as the groups of vertices with dense links and strong interactions. Therefore, many local similarity measures should be closely related to the concept of community structures in networks. Because of the positive correlation between the community structures and the local similarity measures, here, 12 typical similarity measures based on local topological structures are used to extract the local structural information that is related to community structures in networks, and redefine the weight of existing edges in the networks (See Table 1 for the definitions of the local similarity measures).

(1) Common-Neighbor similarity measure (CN) [36]: CN denotes the number of common neighbors between two vertices. In recent years, this measure has widely been applied to the link prediction in complex networks. In fact, two vertices sharing many common neighbors may have more common features, and thus they are more likely to belong to the same grouping and have the stronger interaction and larger weight of edge. Several local similarity measures are based on the number of common neighbors, yet with different normalization methods.

(2) Salton similarity measure (Salton) [37]: It is defined as the number of common neighbors dividing by $\sqrt{k_x \times k_y}$, where $k_x$ is the degree of node $x$. Salton is also called the cosine similarity in the some literatures.

(3) Jaccard similarity measure (Jaccard) [38]: It is defined as the number of common neighbors divided by the *sum* of the number of two vertices' neighbors, which was proposed by Jaccard originally in the context of sets while not networks (it is equal to the ratio of the cardinality of the intersection of two vertices' neighbors sets to the cardinality of the set union).

(4) Sørensen similarity measure (Sørensen) [39]: It is defined as the number of common neighbors dividing by the *average* of two vertices' degrees, which is used mainly for ecological community data.

(5) Hub-Promoted similarity measure (HP) [40]: It is defined as the number of common neighbors dividing by the *minimum* of the number of two vertices' neighbors, which is proposed for quantifying the topological overlap of pairs of substrates in metabolic networks. Under this measurement, the links adjacent to hubs are likely to be assigned high scores since the denominator is determined by the lower degree only.

(6) Hub-Depressed similarity measure (HD) [36]: It is defined as the number of common neighbors dividing by the *maximum* of the number of two vertices' neighbors. Analogously to HP, it considers the opposite effect on hubs.

(7) Leicht-Holme-Newman similarity measure (LHN) [41]. It assigns high similarity to vertex pairs that have many common neighbors compared not to the possible maximum, but to the expected number of such neighbors. Here (see Table 1), the denominator is proportional to the expected number of common neighbors of two vertices in the configuration model (a random graph model which has the same degree distribution as the network under study).

(8) Preferential-Attachment similarity measure (PA) [36]: The preferential attachment similarity obtained discussion, due to the two possible reasons: (a) it requires the least information since it only depends on the degrees of related vertices; (b) it is originated from the popular preferential attachment evolving mechanism of generating the scale-free network, where the probability that this new link will connect two vertices is proportional to the product of the degrees of the two vertices.

(9) Adamic-Adar similarity measure (AA) [42]: It refines the simple counting of common neighbors by assigning the less-connected neighbors more weight (See Table 1).

(10) Resource-Allocation similarity measure (RA) [36]: Motivated by the resource allocation process taking place on networks, Zhou et al. designed the resource allocation similarity to refine the simple counting of common neighbors (see Table 1). In fact, RA is not only closely related to the resource allocation process, but also can successfully suppress the contribution of common neighbors with high degree by assigning them less weight.

(11) Local-Path similarity measure (LP) [36, 43]: It takes consideration of local paths, with wider horizon than CN (see Table 1). This measure reduces to CN if $\varepsilon=0$. $(A^2)_{xy}$ is the number of common neighbors of vertices $x$ and $y$, which is equal to the number of different paths with length 2 connecting $x$ and $y$, and if $x$ and $y$ are not directly connected, $(A^3)_{xy}$ is equal to the number of different paths with length 3 connecting $x$ and $y$. Although it needs more information than CN, it is still a local measure with relatively lower complexity.

(12) Lai-Lu-Nardini similarity measure (LLN) [33]: In general, the dynamic processes triggered on vertices in the same community possess similar behavior patterns, but dissimilar on vertices in different communities. Based on the observation, Lai et al. proposed the similarity measure based on local random walk dynamics. The probability of a walker from one vertex to another in $t$-step random walk is determined by matrix $P^t$ ($t$ is random walk length, determining the range of the local structure that will be browsed). In general networks, good results can often be obtained by using a small $t$-value ($t$=2, 3, …). The element $p_{xy}$ of the transition matrix $P$ is the ratio between the weight of edge $(x, y)$ and the weighted degree of vertex $x$, *i.e.* $p_{xy} = w_{xy}/w_x$. The behavior patterns of the random walk dynamics from each vertices can be quantified by a N-dimensional vector $v_x$ ($x$=1 to $N$)—the row of the matrix $\sum_{\tau=1}^{t} p^{\tau}$ where all of random walks whose steps vary from 1 to $t$ are taken into account to reinforce the contributions from the vertices near the target vertices currently considered. See Table 1 for the definition of *LLN*. If the behavior vectors $v_x$ and $v_y$ are highly consistent, then $S_{xy}^{LLN} \to 1$, 0 otherwise. Generally, the pairs of vertices in the same communities have higher values of the similarity than those between communities.

Note that, in order to avoid the edge weight of the value of zero between vertices without common neighbors, the number of common neighbors between two vertices adds a value of 1 for CN, Salton, Jacard, Sørensen, HP, HD, LHN as well as LP, while the original definitions of AA and RA will add a factor of $1/\log(k_{max})$ and a factor of $1/k_{max}$ respectively, which should not affect markedly the original definitions.

**Table 1:** The definition of the local similarity measures. $\Gamma(x)$ and $\Gamma(y)$ denote the sets of neighbors of vertices $x$ and $y$ respectively; |*| is the cardinality of the set *; $k_x$ and $k_y$ denote the degrees of vertex $x$ and vertex $y$ respectively; $v_x$ and $v_y$ are the behavior vectors of vertices $x$ and $y$ in the random walks, while ($v_x$, $v_y$) denotes the dot product of the two vectors.

| Name | Definition | Name | Definition |
|---|---|---|---|
| CN | $s_{xy}^{CN} = \left\| \Gamma\ x\ \cap \Gamma\ y\ \right\|$ | LHN | $s_{xy}^{LHN} = \dfrac{\left\| \Gamma\ x\ \cap \Gamma\ y\ \right\|}{k_x \times k_y}$ |
| Salton | $s_{xy}^{Salton} = \dfrac{\left\| \Gamma\ x\ \cap \Gamma\ y\ \right\|}{\sqrt{k_x \times k_y}}$ | PA | $s_{xy}^{PA} = k_x \times k_y$ |
| Jaccard | $s_{xy}^{Jaccard} = \dfrac{\left\| \Gamma\ x\ \cap \Gamma\ y\ \right\|}{\left\| \Gamma\ x\ \cup \Gamma\ y\ \right\|}$ | AA | $s_{xy}^{AA} = \sum_{z \in \Gamma\ x\ \cap \Gamma\ y} \dfrac{1}{\log k(z)}$ |
| Sørensen | $s_{xy}^{Sørensen} = \dfrac{\left\| \Gamma\ x\ \cap \Gamma\ y\ \right\|}{(k_x + k_y)/2}$ | RA | $s_{xy}^{RA} = \sum_{z \in \Gamma\ x\ \cap \Gamma\ y} \dfrac{1}{k(z)}$ |
| HP | $s_{xy}^{HP} = \dfrac{\left\| \Gamma\ x\ \cap \Gamma\ y\ \right\|}{\min\{k_x, k_y\}}$ | LP | $s_{xy}^{LP} = (A^2 + \varepsilon A^3)_{xy}$ |
| HD | $s_{xy}^{HD} = \dfrac{\left\| \Gamma\ x\ \cap \Gamma\ y\ \right\|}{\max\{k_x, k_y\}}$ | LLN | $s_{xy}^{LLN} = \dfrac{(v_x, v_y)}{\sqrt{(v_x, v_x)}\sqrt{(v_y, v_y)}}$ |

## 2.2 Application of local similarity measures to community detection

Here, several classical modularity-based community detection methods are analyzed: Girvan-Newman algorithm (GN) — a typical divisive algorithm based on edge betweenness [44], which generally used the modularity as the stopping principle and is also a popular reference of many community detection methods; Newman's fast greedy algorithm (NF) — a typical agglomerative algorithm based on greedy optimization of modularity, which has low computational complexity but low accuracy for searching the optimal modularity [45]; (3) Fast algorithm of Blondel et al (BF) based on modularity optimization, which has low computational complexity and very high accuracy for searching the optimal modularity [25]. Moreover, we also test several community detection methods based on different strategies: Infomap [46], which uses random walks for information flows and detects community structure by compressing them; COPRA [47], which is based on the label propagation concept in [13] where each vertex is assigned initially a unique label, then the labels propagate by vertex neighbors and finally the vertices in each community are to have the same label after several iterations; OSLOM [48], which is based on the local optimization of a fitness function describing the statistical significance of communities with respect to random fluctuations.

The local similarity measures are applied to the community detection in networks by the following procedure:

Step 1: Calculate the local similarity measure between vertices that are connected by edges in a network;

Step 2: Redefine the weight of the existing edges in the network by the local similarity measure between the ends of the edges;

Step 3: Feed the finally weighted network into the corresponding community detection algorithm, and output the community division of the network.

This procedure can be divided into two main phases: preprocessing the original network (step 1-2) and dividing the newly obtained network into communities (step 3). Because the extraction of the local information has low computational complexity, the complexity of the procedure depends on the community detection algorithm itself.

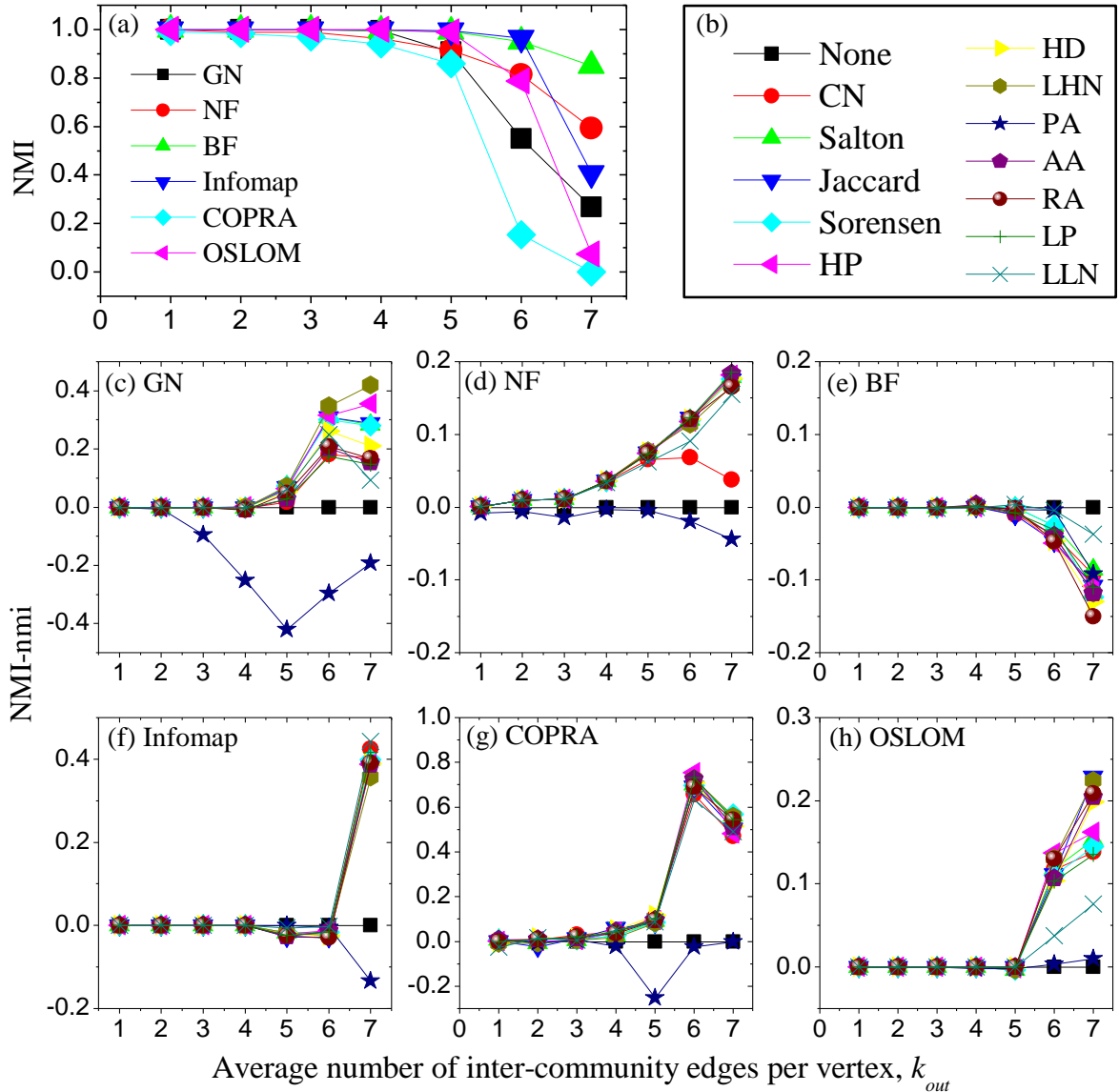

**Figure 1.** (a) The *normalized mutual information* (NMI) [21] between the real and found communities in the Girvan-Newman networks as a function of the average number $k_{out}$ of inter-community edges per vertex, obtained by the original methods. (b) Each symbol denotes the combination of the similarity measure with the corresponding methods in (c)-(h), that is, to preprocess the original networks firstly by the similarity measure and then use the corresponding methods to discover the community structures in the networks; 'None' denotes the results of the original methods themselves. (c)-(h) The increment of the *normalized mutual information* (i.e. NMI-nmi) after the introduction of the similarity measure for each method. In the following figures 1, 4 and 7, the symbols denote the same meanings.

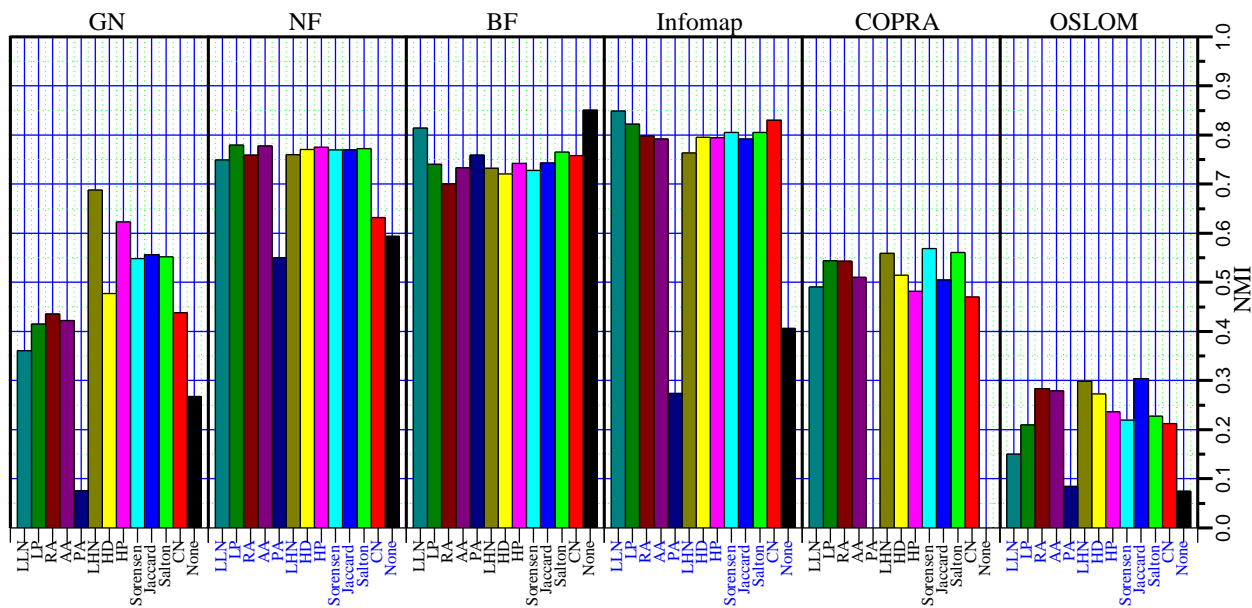

**Figure 2.** Each bar denotes the NMI of the corresponding community detection methods with the corresponding similarity measure in the Girvan-Newman networks with $k_{out}$= 7. 'None' denotes the results of the original methods themselves.

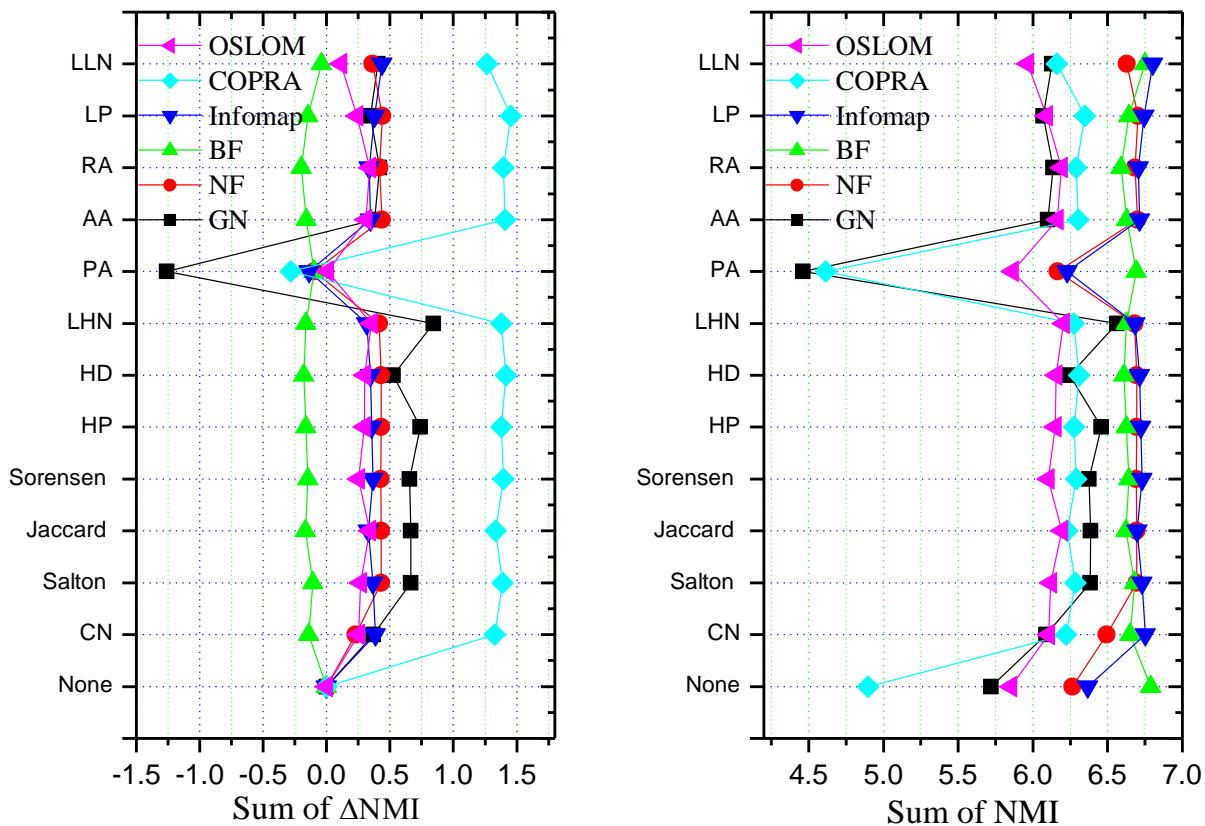

**Figure 3.** Comparison of overall performance: (Left) the sum of ∆NMI (i.e. NMI-nmi) and (Right) the sum of NMI, in all the Girvan-Newman networks with $k_{out} \leq 7$, for each community detection method with different similarity measures.

## 3. Experimental results

### 3.1. Girvan-Newman networks

The Girvan-Newman networks [44], a type of representative homogeneous networks with community structures, consist of four communities of 32 vertices. Each vertex in the networks has the total expected degree $k_{in}+k_{out}=16$, where $k_{in}$ is the average number of edges connecting it to vertices in the same community and $k_{out}$ is the average number of edges to vertices in other communities. With $k_{out}$ increasing from zero, the communities in the networks become more and more difficult to identify. In figure 1, we show the performance of different community detection methods in the networks and the effect of the local similarity measure on each method, as a function of the average number of inter-community edges per vertex in the networks.

It is not our purpose to compare the methods themselves, while figure 1(a) provides the performance comparison of the original methods in the test networks. The methods can give good results in the networks. To clearly display the effect of the local structural similarity measures on community detection methods, we calculate the increment (NMI-nmi) of the normalized mutual information [21] for each method caused by the introduction of the local structural similarity measures, where NMI and nmi denote the normalized mutual information obtained respectively by the corresponding method with the similarity measure and by the original method. The values of NMI-nmi>0 denote that the local structural similarity measure has the positive effect on the methods; NMI-nmi<0 means a negative effect; while NMI-nmi=0 means no effect. As shown in figures 1(c)-(h), on the whole, the local information extracted by most of the local similarity measures can indeed help in improving the community detection in the networks. However, PA is an exception. The addition of PA cannot generate better results for most methods in the networks. Especially, PA badly weakens the performance of GN even for small values of $k_{out}$. Maybe, PA itself cannot provide enough and useful information for community structures to the methods.

The similarity measures of intra-community edges generally have larger values than the inter-community edges in networks. So they may help GN in identifying the inter-community edges in the networks. For GN, these similarity measures indeed bring better results except for PA. The top improvement of performance is obtained by LHN, HP, Jaccard, Sorensen and Salton (see figure 1(c) and figure 2 for GN). Especially LHN seems to be very excellent for GN.

For NF, the similarity measures can help the greedy optimization procedure of NF to agglomerate vertices into communities more exactly. Therefore, similarly, the similarity measures, except for PA, can clearly improve results of NF in the networks (see figure 1(d) and figure 2 for NF). As we see, most of the similarity measures have similar behaviors. Moreover, compared to other similarity measures, CN can only bring a few improvements in the networks, though it is the base of several similarity measures (e.g. Salton and Jacard).

Unexpectedly, none of the similarity measures shows the improvement for BF. They even reduce the performance of BF obviously in the networks. This may be because BF itself has had very excellent performance in the networks; there is less room for improvement, while the similarity measures may disturb the optimization procedure of BF itself (note that the disturbance of LLN is the weakest). However, GN and NF still have large rooms for improvement in the networks, so the local similarity measures can help improve the performance of them.

For Infomap, the behaviors of the similarity measures is similar in community detection, while the similarity measures seem to go against Infomap when $k_{out}=5$ and 6 (see figure 1(f) for Infomap). This may be because the original Infomap is being at unstable status and the performance of it also begins to decline in the cases (figure 1(a)). When $k_{out}=7$, Infomap itself becomes very poor (NMI is only 0.4), while the similarity measures are able to greatly enhance the performance of it, and LLN seems to give best results, followed by LP and CN (figure 2 for Infomap).

For COPRA, the similarity measures also show the similar behaviors in the networks, and have a slight disturbance for small $k_{out}$-values (see figure 1(g)). This may be also because COPRA is unstable in the cases (figure 1(a)). With the increase of $k_{out}$(≥3), the improvement becomes more and more obvious. And the excellent performance for $k_{out}=7$ seems to be given by LHN, Sorenson and CN, followed by LP and RA (figure 2 for COPRA).

For OSLOM, the similarity measures can generate obvious improvement when $k_{out}$>5. In the networks with $k_{out}=7$, the similarity measures can be classified into four sets: (Jaccard, LHN, RA, AA and HD), (LP, HP, Sorenson, Salton and CN), (LLN) and (PA). LLN is clearly weaker than the first and second sets. Interestingly, PA seems to be able to generate slight improvement too.

Finally, we display the comprehensive effect of the local structural similarity measures on different methods. Figure 3 shows the cumulative improvement of performance and the cumulative performance of the methods. We can find that, for all the local structural similarity measures, the largest cumulative improvement is obtained by COPRA, while the smallest cumulative improvement is obtained by BF (negative effect). The best cumulative performance is obtained by Infomap with LLN as well as the original BF.

### 3.2. Lancichinetti-Fortunato-Rachicchi (LFR) networks

Differently from the above Girvan-Newman networks, the LFR networks [49] consider the heterogeneity of vertex degrees and community sizes in realistic networks. The power-law distributions of vertex degrees and community sizes in the networks are controlled by the exponents $\gamma$ and $\beta$ respectively, and the common mixing parameter $\mu$ controls the ratio between the external degree of each vertex with respect to its community and the total degree of the vertex. Other parameters of the networks: $N$ is the number of vertices, $<k>$ and $k_{max}$ are the average and maximum degree, $C_{min}$ and $C_{max}$ are the minimum and maximum community sizes. Here, we will construct two sets of the LFR networks with:

(1) moderate heterogeneity of community sizes ($\gamma = -2$, $\beta = -2$, $C_{min}$=20 and $C_{max}$=50) and

(2) strong heterogeneity of community sizes ($\gamma = -2$, $\beta = -2$, $C_{min}$=10 and $C_{max}$=100).

Other parameters: $N$=1000, $<k>$=20, and $k_{max}$ =50.

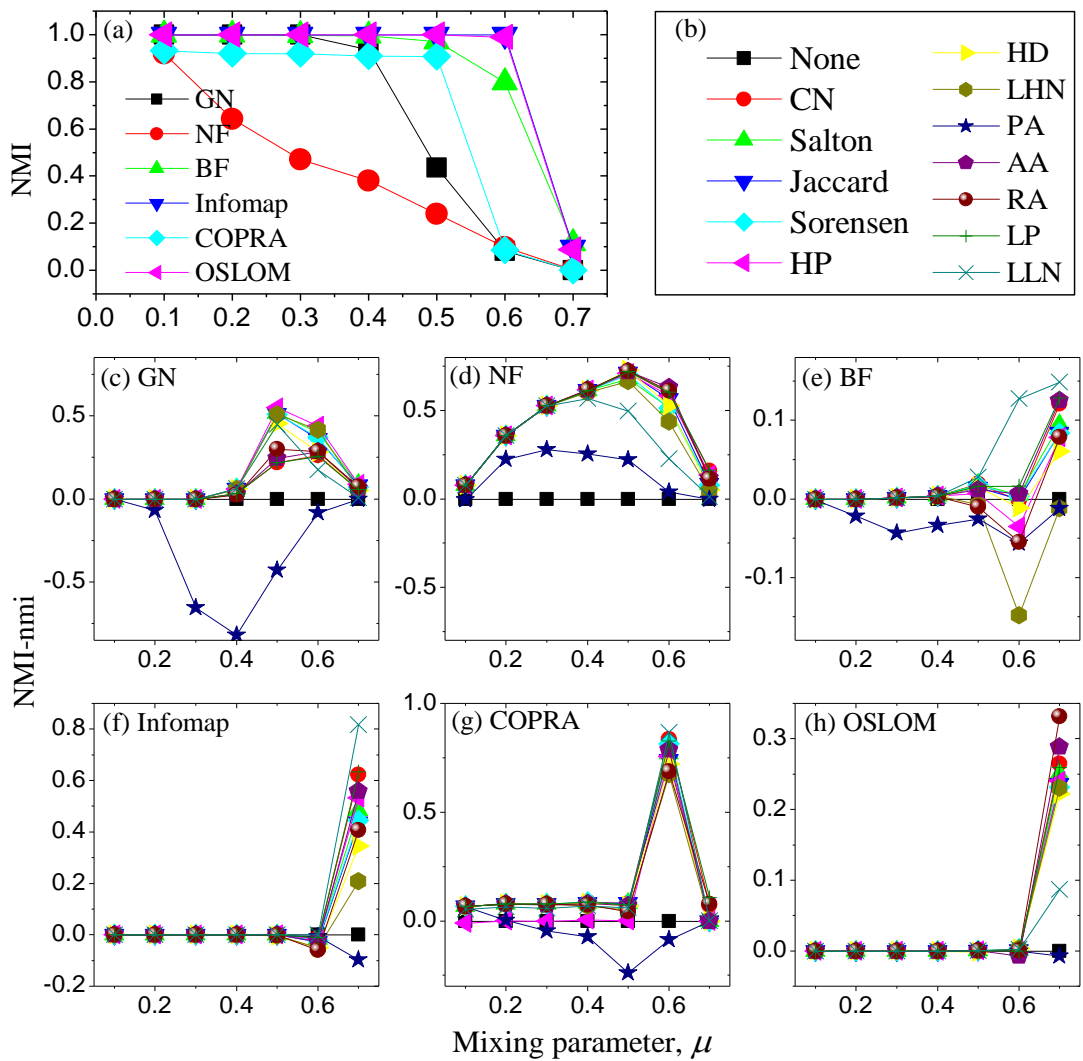

**Figure 4.** (a) The *normalized mutual information* (NMI) between the real and found communities, in the LFR networks with moderate heterogeneity of community sizes, as a function of the mixing parameter $\mu$, obtained by the original methods. (b)-(h) The meanings of other symbols are the same as in figure 1.

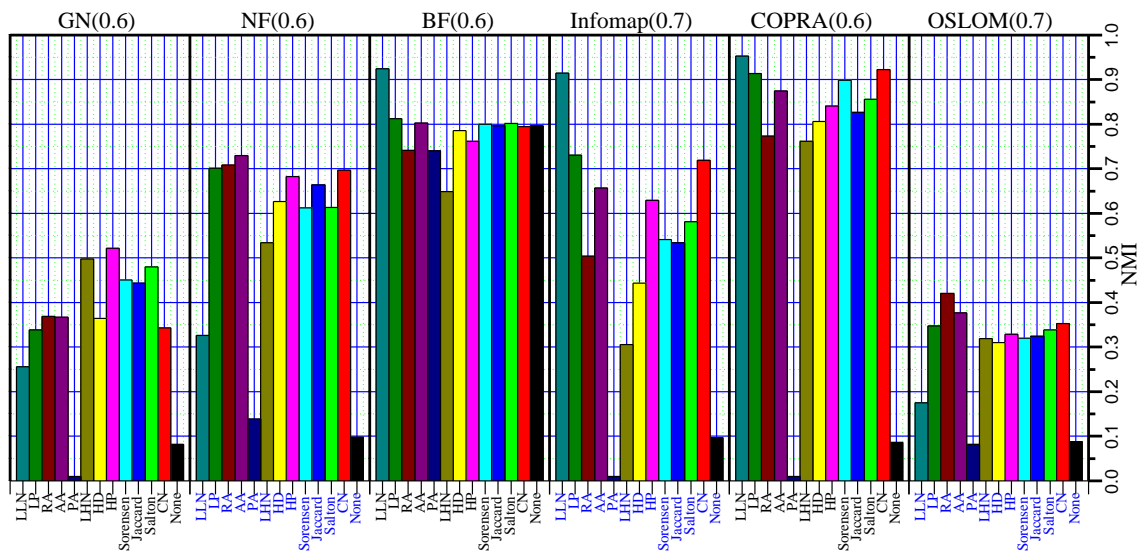

**Figure 5.** The bars denote the NMI in the LFR networks with moderate heterogeneity of community sizes, obtained by GN, NF, BF and COPRA ($\mu$=0.6) and by Infomap and OSLOM ($\mu$=0.7).

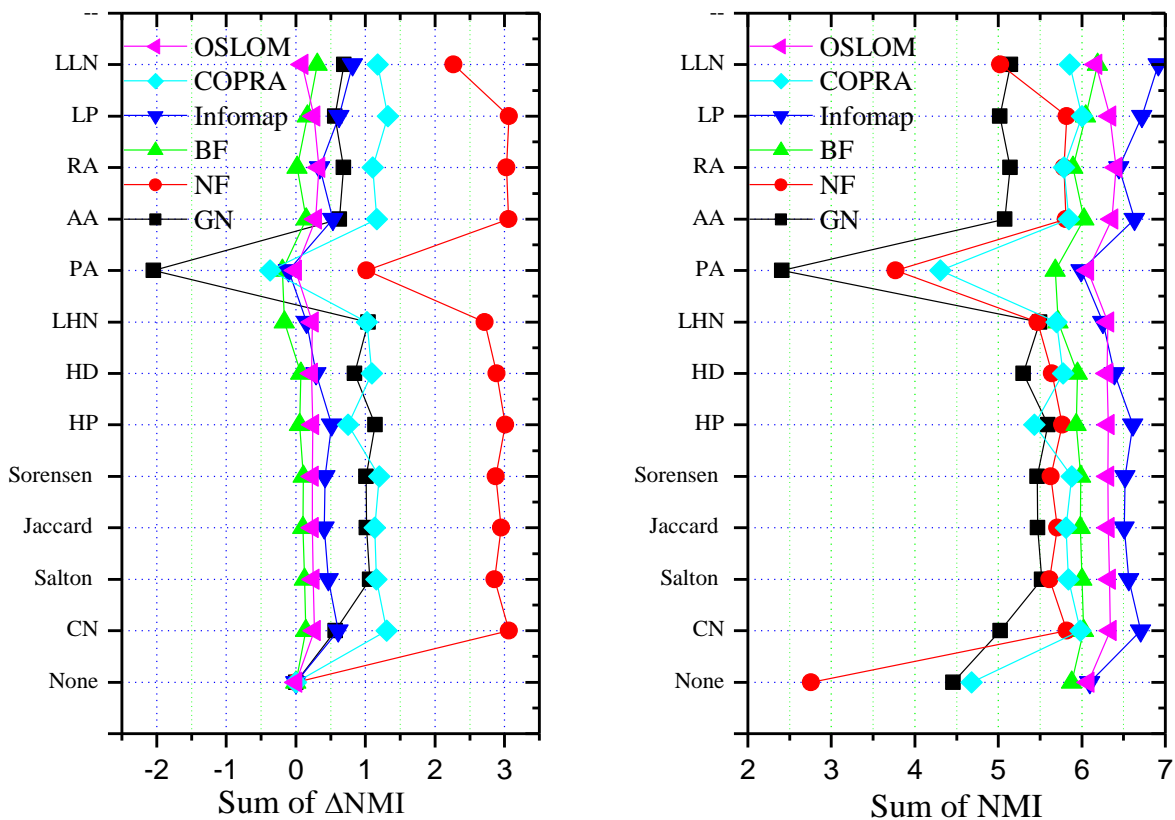

**Figure 6.** Comparison of overall performance: (Left) the sum of $\Delta$NMI (i.e. NMI-nmi) and (Right) the sum of NMI, in all the LFR networks with moderate heterogeneity of community sizes and $\mu \leq 0.7$, for each community detection method with different similarity measures.

### 3.2.1. LFR networks with moderate heterogeneity of community sizes

In figure 4, we show the performance of different methods in the LFR networks and the effect of the local similarity measures on different methods, as a function of the mixing parameter $\mu$. With the increase of $\mu$, the communities in the networks become more and more difficult to identify. For the original methods, the best performance is given by Infomap and OSLOM, followed by BF, while the improvements resulting from the similarity measures are closely related to the original methods (figure 4). The improvement of BF, Infomap and OSLOM is not clear, until $\mu$=0.7. This may be because they have excellent performance in the networks and the local similarity measures cannot give the methods enough reliable information about the community structures. For all the methods, PA still brings the poorest results in the networks.

For different methods, the excellent performance is obtained by different local similarity measures. In Figure 5, the excellent local similarity measures are, respectively, LHN, HP, Sorenson, Jaccard and Salton for GN (NMI>0.4 for $\mu$=0.6), LP, RA, AA, HP, Jaccard and CN for NF (NMI>0.65 for $\mu$=0.6), LLN and LP for BF, LLN, LP and CN for Infomap (NMI>0.7 for $\mu$=0.7), LLN, LP and CN for COPRA (NMI>0.9 for $\mu$=0.6), as well as RA, AA, LP and CN for OSLOM (NMI≥0.35 for $\mu$=0.7). Unexpectedly, PA also shows the slight improvement for NF.

Figure 6 (Left) shows the largest cumulative increment of performance is obtained by NF. NF itself is the poorest in the networks. It cannot correctly find the predefined community structures, even for the small $\mu$-values. The main reason is not due to the resolution limit of modularity but the very weak ability of searching community structures especially in the large-size networks. So it leaves large room for improvement. As expected, the results show that the effect of the similarity measures on NF is extraordinary. By using the similarity measures, NF can generate similar or even better results than GN in the networks. On the whole, the local similarity measures can help in improving the performance of the methods in the networks. For the cumulative performance (figure 6 (Right)), Infomap seems to generate the best results (especially for LLN) while GN is the poorest on the whole.

### 3.2.2. LFR networks with strong heterogeneity of community sizes

In the LFR networks with strong heterogeneity of community sizes, one of typical problems is that the resolution limit of modularity becomes more serious (see figure 7). Even if $\mu$ is very small, the modularity optimization cannot also identify all embedded communities in the networks. Both of NF and BF are very poor in the networks, though BF itself has excellent performance for searching optimal modularity. GN only partly depends on the modularity, so it still has good results until $\mu$>0.4. In these LFR networks, GN, NF BF as well as COPRA have the great rooms for improvement. While Infomap and OSLOM still have excellent performance in the networks, which means that the resolution limit of modularity does not seem to affect the two methods.

As expected, the local similarity measures indeed bring the great improvement for GN, NF, BF and COPRA (figure 7 and figure 8), while, for Infomap and OSLOM, the behaviors of the local similarity measures are similar to those in figure 4. Of course, PA still brings the poorest results. For small $\mu$-values (e.g. $\mu$=0.2 and 0.3), modularity optimization (NF and BF) cannot identify the predefined community structures due to the resolution limit problem of modularity. In this case, the problem can be solved completely by the similarity measures, resulting NMI=1 for NF and BF (including GN). With the increase of $\mu$, the resolution limit of modularity becomes more and more serious (because the connections between communities increase). And the community structures also become more and more indistinct, so the search for the community structures in the networks (or the optimization of modularity) becomes more and more difficult. As a result, for example, when $\mu$=0.6 or 0.7, NMI<1 for most methods, even with the help of the local similarity measures (figure 8). But the effect of the similarity measures is still positive.

There are several possible reasons for the positive effect: (1) the increase of the identifiability of communities, (2) the enhancement of the search procedure for optimal modularity (especially for the methods that the ability of searching optimal modularity is not well, like NF) and (3) the improvement of the resolution limit of modularity (for all modularity-based methods). This main reason for BF is the mitigation of the resolution limit of modularity, because the optimal search for BF, as shown in the above section, cannot obviously been improved. For other methods, the main reason may be that the local similarity measures make the community structures more obvious, increasing the identifiability of communities.

In the LFR networks, the excellent performance of different methods is also obtained by different local similarity measures (figure 8): e.g. LHN, HP, Sorenson, Jaccard and Salton for GN (NMI>0.8); LP, RA, AA, Jaccard and CN for NF (NMI>0.8); all similarity measures for BF (NMI>0.8, except for PA); LLN, LP, AA and CN for Infomap (NMI>0.7); LLN, LP, AA, Sorenson and CN for COPRA; and LP, RA, AA and Salton for OSLOM (NMI>0.45).

Finally, we examine the cumulative effect of the local similarity measures on the methods (figure 9). The largest cumulative improvement of the performance is still obtained by NF. For Infomap and COPRA, the local similarity measures can generate a great increment of NMI for large $\mu$-values, so they have no large difference in the cumulative increment. For the cumulative performance, Infomap still gives the best results (especially for LLN).

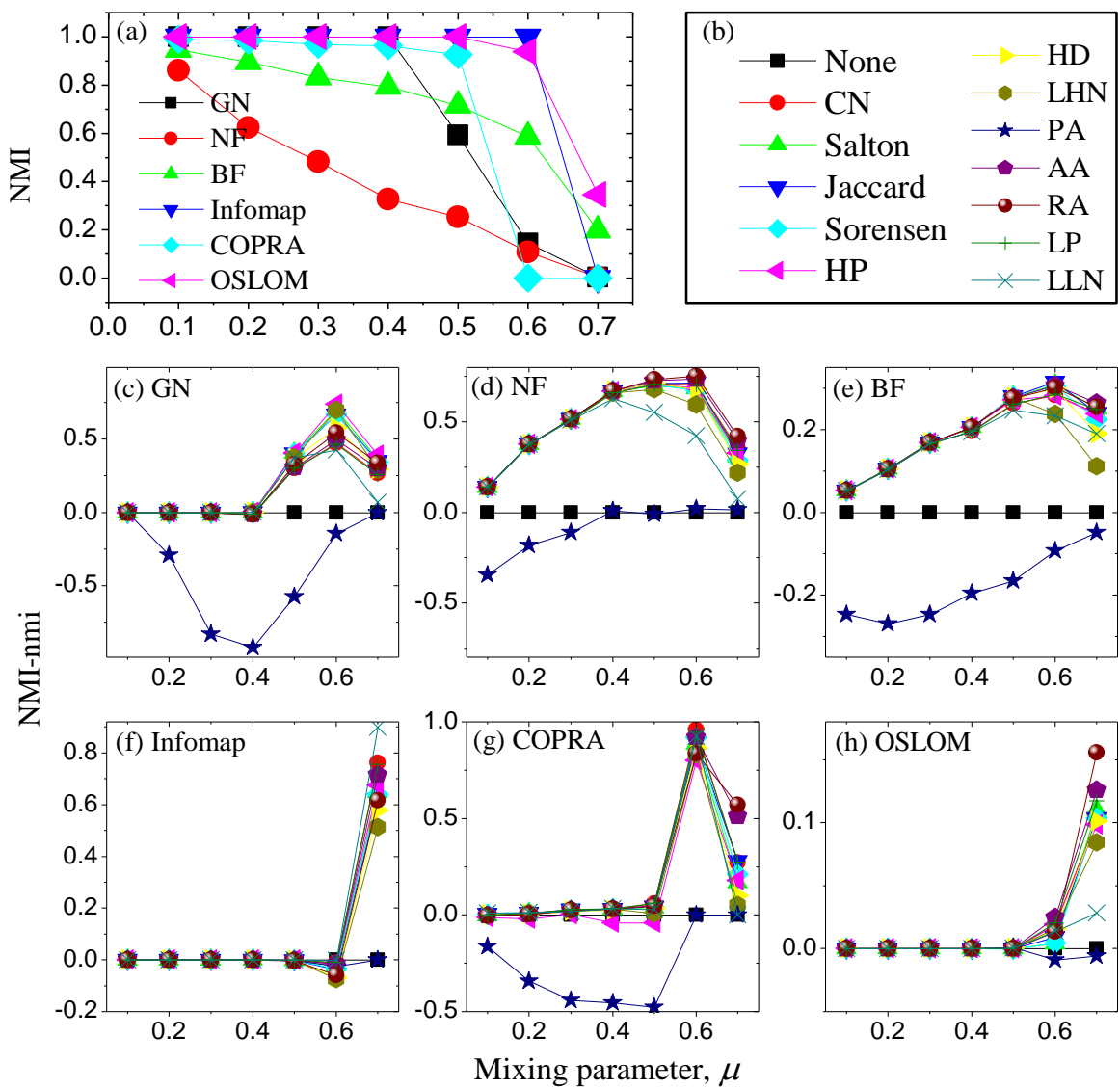

**Figure 7.** (a) The *normalized mutual information* (NMI) between the real and found communities, in the LFR networks with *strong* heterogeneity of community sizes, as a function of the mixing parameter $\mu$, obtained by the original methods. (b)-(h) The meanings of other symbols are the same as in figure 1.

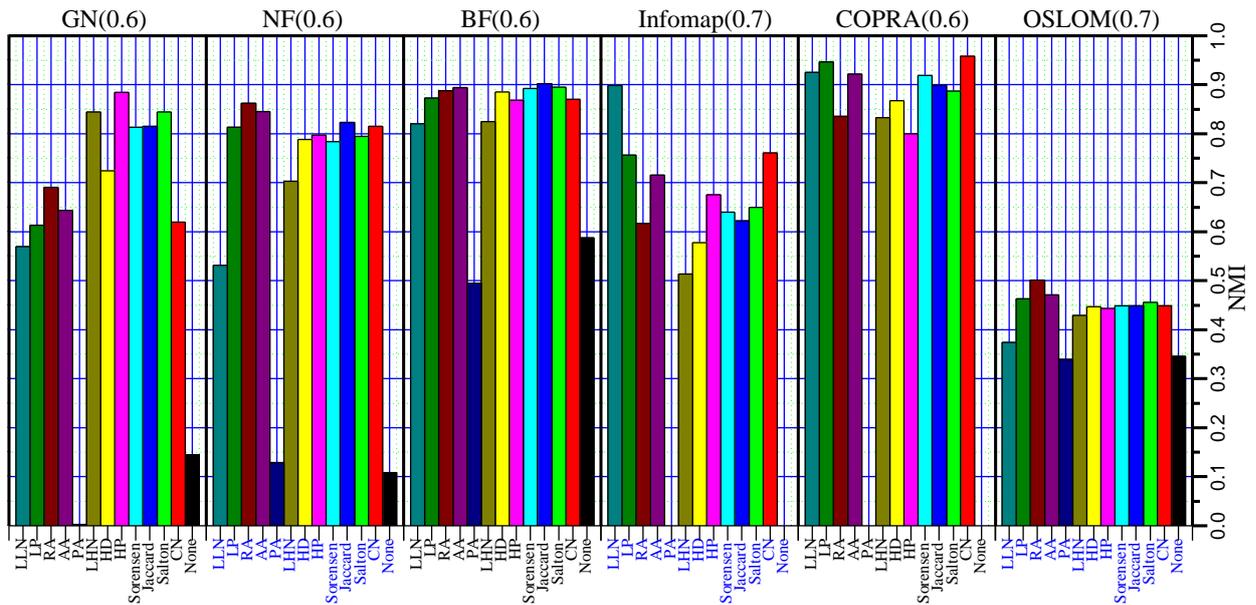

**Figure 8.** The bars denote the NMI in the LFR networks with *strong* heterogeneity of community sizes, obtained by GN, NF, BF and COPRA ($\mu$=0.6) and by Infomap and OSLOM ($\mu$=0.7).

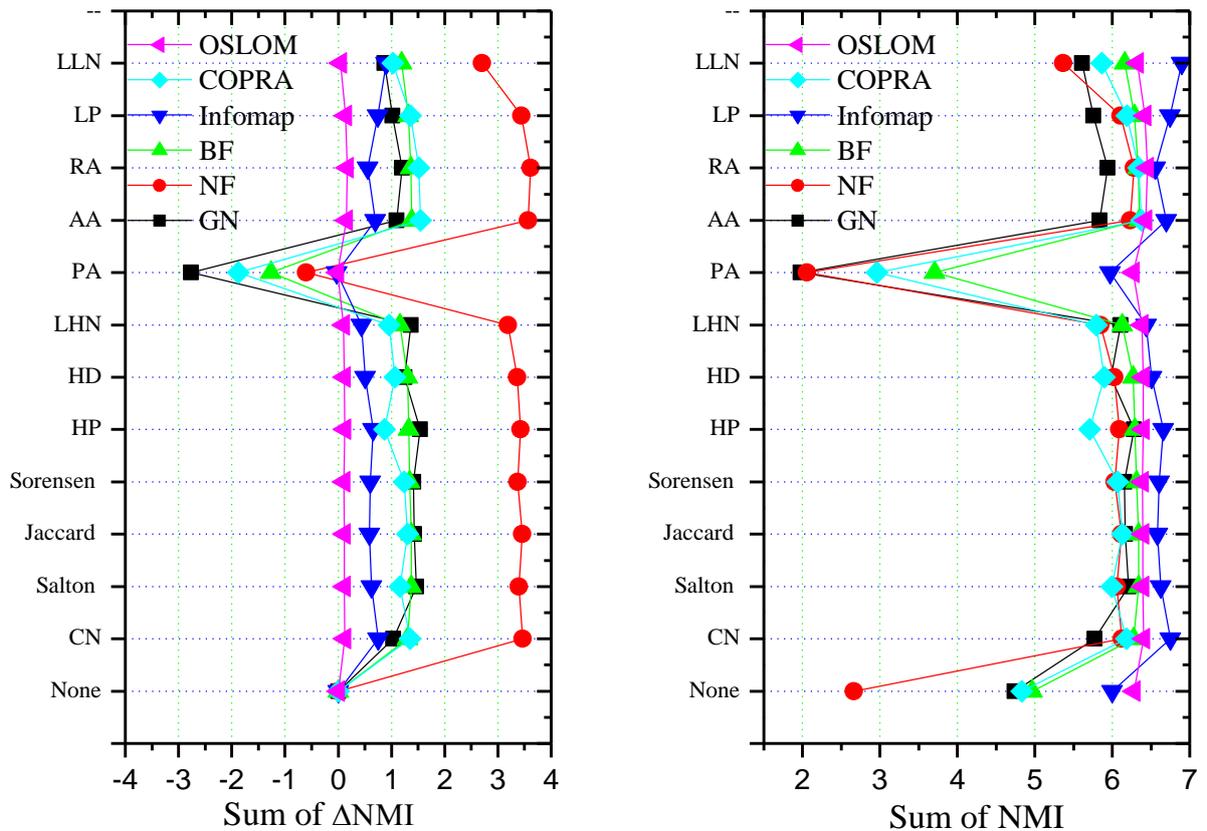

**Figure 9.** Comparison of overall performance: (Left) the sum of $\Delta$NMI (i.e. NMI-nmi) and (Right) the sum of NMI, in all the LFR networks with strong heterogeneity of community sizes and $\mu \leq 0.7$, for each community detection method with different similarity measures.

### 3.2.3. Mitigating the resolution limit of modularity

As shown above, the local similarity measures are helpful for the resolution limit problem of modularity, but cannot solve the problem thoroughly. Here, we further discuss the problem in a set of the regular networks and a set of the general networks by gradually increasing the network size (see figure 7). The networks have clear community structures, but modularity optimization (as well as other methods that depend on modularity) is unable to correctly identify the embedded community structures in the networks when the network size is very large, due to the resolution limit problem of modularity.

The appropriate edge-reweighting could strengthen the weight of the intra-community edges, while weaken the weight of the inter-community edges, so it will be able to enhance the ability of the modularity-based methods to put up with the resolution problem of modularity. As shown in figure 10, the local similarity measures can effectively help solve the resolution limit problem of modularity in the *small-size* networks, but they only are able to mitigate the problem in the *large-size* networks (because the problem still exist). In this case, one can consider to repeatedly weighting the edges in the networks by the local similarity measures, where the definitions of the similarity measures need to extend to the weighted case). However, because of the lack of the information, it is not easy to determine how many times the edge-weighting process should be repeated. Which similarity measure is the best choice for mitigating the resolution limit of modularity? This seems related to the networks under study. While we note that LLN (as well as CN and LP) seems not to be good at dealing with the resolution limit problem of modularity in the networks; PA makes the problem of modularity optimization worse.

Moreover, GN itself does not depend on modularity, and thus it is not affected by the resolution limit problem of modularity. In the above networks, GN is able to produce a clear dendrogram that contains all the groups, while it needs reasonable stopping criteria to determine the final community division as output. If modularity is used as the stopping criterion, then the resolution limit problem of modularity will still appear. To avoid the problem, one can consider to make use of other quality functions or principles of communities, such as the local modularity [50] and the strong or weak definitions of communities [51].

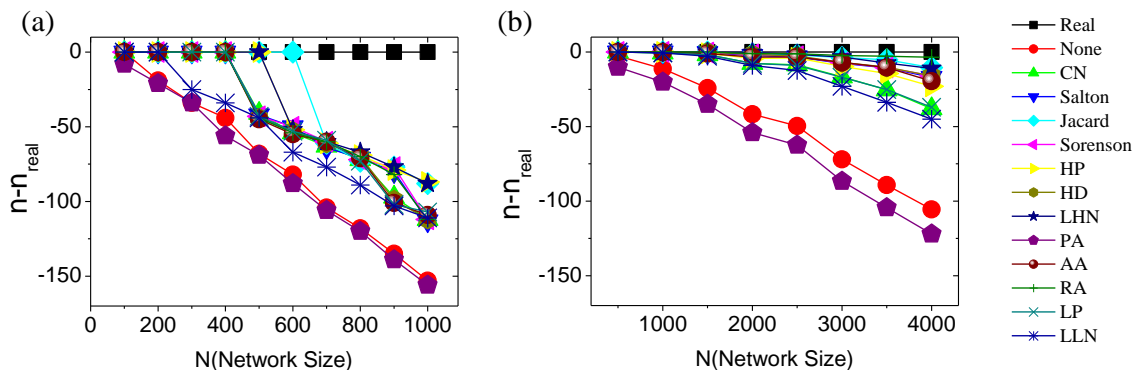

**Figure 10.** Effect of local similarity measures on the resolution limit of modularity. $n$-$n_{real}$ denotes the difference between the number of the identified groups and the real groups. When $n$-$n_{real}$ =0, the predefined community structures are identified; when $n$-$n_{real}$ <0, there exists the merger of the predefined groups in the networks under study. (a) The networks consisting of cliques with 5 vertices, where the network size ($N$) increases by increasing the number of the cliques. (b) The LFR networks with strong community-size heterogeneity ($\gamma$=-2, $\beta$=-2, $\mu$=0.4, $C_{min}$=10, $C_{max}$=100, $<k>$=20, and $k_{max}$ =50, while $N$ will vary). The two kinds of networks both have clear community structures, while the resolution limit problem of modularity will become more and more serious with the increase of the network size (here, BF is used, due to its low computational complexity and high accuracy for modularity optimization).

### 3.3. Real-world networks

Finally, we apply the above methods to the real-word networks: Football [44], Jazz [52], Email [53] and Yeast [54]. For convenience of quantitative examination, we calculate the modularity obtained by the above methods on the real-world networks (figure 11). Clearly, the modularity in the networks obtained by almost all the methods can increase due to the introduction of the local similarity measures, except for PA. The results show that the local similarity measures can lead to obvious improvement for the methods, if the modularity is used as the evaluation criterion for the community quality (though it is not the best choices). The best local similarity measures may be different for different methods and different networks, while LHN seems to be able to bring relatively higher modularity for the methods in the above networks.

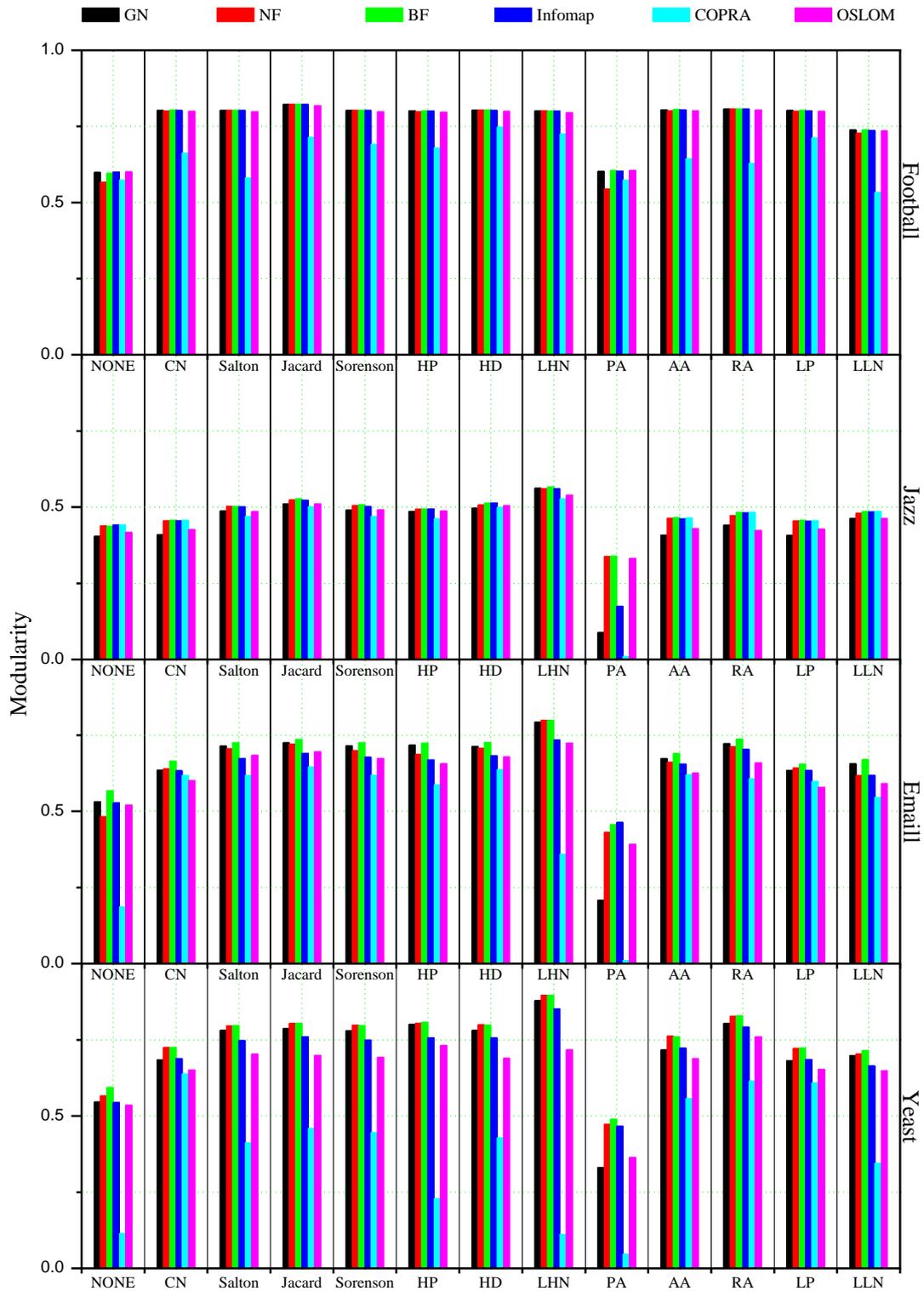

**Figure 11.** Modularity in real-world networks, obtained by different community detection methods with different structural similarity measures.

**Table 2:** Comparison of cumulative increment of performance (i.e. Sum of NMI-nmi) over all the above test networks, for all the combinations of the methods with the similarity measures.

| Similarity | GN | NF | BF | Infomap | COPRA | OSLOM |
|---|---|---|---|---|---|---|
| None | 0.00 | 0.00 | 0.00 | 0.00 | 0.00 | 0.00 |
| CN | 1.97 | 6.75 | 1.31 | 1.75 | 3.98 | 0.64 |
| Salton | 3.19 | 6.69 | 1.39 | 1.46 | 3.71 | 0.63 |
| Jaccard | 3.11 | 6.84 | 1.31 | 1.32 | 3.76 | 0.69 |
| Sorensen | 3.08 | 6.67 | 1.31 | 1.39 | 3.83 | 0.59 |
| HP | 3.41 | 6.87 | 1.22 | 1.53 | 3.00 | 0.66 |
| HD | 2.61 | 6.68 | 1.19 | 1.15 | 3.57 | 0.64 |
| LHN | 3.25 | 6.32 | 0.83 | 0.91 | 3.36 | 0.69 |
| PA | -6.07 | 0.31 | -1.55 | -0.27 | -2.53 | -0.01 |
| AA | 2.10 | 7.06 | 1.37 | 1.58 | 4.12 | 0.74 |
| RA | 2.30 | 7.07 | 1.18 | 1.24 | 4.01 | 0.84 |
| LP | 1.93 | 6.95 | 1.33 | 1.75 | 4.13 | 0.64 |
| LLN | 1.97 | 5.34 | 1.46 | 2.15 | 3.47 | 0.25 |

**Table 3**: Comparison of cumulative performance (i.e. Sum of NMI) over all the above test networks, for all the combinations of the methods with the similarity measures.

| Similarity | GN | NF | BF | Infomap | COPRA | OSLOM |
|---|---|---|---|---|---|---|
| None | 14.92 | 11.68 | 17.64 | 18.46 | 14.41 | 18.21 |
| CN | 16.89 | 18.43 | 18.95 | 20.21 | 18.38 | 18.85 |
| Salton | 18.11 | 18.37 | 19.02 | 19.92 | 18.12 | 18.85 |
| Jaccard | 18.02 | 18.52 | 18.94 | 19.79 | 18.17 | 18.90 |
| Sorensen | 18.00 | 18.35 | 18.94 | 19.86 | 18.24 | 18.81 |
| HP | 18.33 | 18.55 | 18.85 | 19.99 | 17.41 | 18.87 |
| HD | 17.53 | 18.36 | 18.82 | 19.61 | 17.98 | 18.85 |
| LHN | 18.17 | 18.00 | 18.47 | 19.37 | 17.76 | 18.90 |
| PA | 8.84 | 11.99 | 16.08 | 18.19 | 11.88 | 18.20 |
| AA | 17.02 | 18.74 | 19.00 | 20.04 | 18.52 | 18.96 |
| RA | 17.22 | 18.75 | 18.82 | 19.70 | 18.42 | 19.05 |
| LP | 16.85 | 18.63 | 18.97 | 20.21 | 18.54 | 18.85 |
| LLN | 16.88 | 17.02 | 19.10 | 20.61 | 17.88 | 18.46 |

## 4. Conclusion and discussion

In general, groups of vertices with similar properties are also to be groups of vertices with denser links and stronger interactions, so many similarity measures may be closely related to the community structures in networks. Because of the positive correlation between the community structures and the similarity measures in networks, the similarity measures may be of help in community detection. We investigate systematically how various local similarity measures affect the community detection in various networks and compare the effect of different local similarity measures on community detection.

On the whole, the application of the local similarity measures can help in identifying the inter-community edges (e.g. for GN), searching the optimal modularity (e.g. for NF), mitigating the resolution limit problem of modularity (e.g. for BF and NF), and more importantly, making the community structures more obvious (e.g. for Infomap, OSLOM and COPRA). Note that the local similarity measures can only mitigate the resolution limit problem, but cannot guarantee to solve the problem thoroughly. To detect the embedded communities in the networks, one may have recourse to the multi-resolution methods such as the multi-resolution modularity method, which can help in identifying the multi-scale community structures in networks.

The test results show that the positive effect of the local similarity measures is closely related to the networks under study and the applied methods themselves. By and large, most of the local similarity measures will lead to considerable improvement for the methods, if the original methods have large rooms for improvement. In Tables 2-3, as a final comprehensive comparison, we have shown the cumulative increment of performance and the cumulative performance over all the above test networks for all the combinations of the methods and the similarity measures. They display the comprehensive effect of different local structural similarity measures on different community detection methods. We believe that they should be useful for the readers, though some analysis has been given in the previous sections, and expect that all the data in the paper can provide the readers much valuable information.

Moreover, the extraction of information based on the local similarity measures has the additional advantage of low computational complexity and it hardly affects the speed of the applied community detection methods, because it is only based on the local topological structures in networks. Additionally, it is possible to extract useful information for community detection by other approaches, even by other community detection methods, which should also be able to improve the community detection in networks. Finally, we hope that the work can help in enriching the knowledge for community detection in networks.

## Acknowledgement

We sincerely appreciate the referee's valuable suggestions. This work has been supported by the construct program of the key discipline in Hunan province, the Scientific Research Fund of Education Department of Hunan Province (Grant Nos. 14C0126, 11B128, 14C0127, 14C0112, 12C0505 and 14B024), the Project of Changsha Medical University (Grant No. KY201517), the Department of Education of Hunan Province (Grant No. 15A023), the Hunan Provincial Natural Science Foundation of China (Grant No. 2015JJ6010), the Hunan Provincial Natural Science Foundation of China (Grant No. 13JJ4045), and the National Natural Science Foundation of China (Grant No. 11404178), and partly by the Doctor Startup Project of Xiangtan University (Grant No. 10QDZ20).

## References

[1]   S. Fortunato, Community detection in graphs, 2010 Physics Reports 486 75.

[2]   P. Chen and S. Redner, Community structure of the physical review citation network, 2010 Journal of Informetrics 4 278.

[3]   S.-H. Zhang, X.-M. Ning, C. Ding, and X.-S. Zhang, Determining modular organization of protein interaction networks by maximizing modularity density, 2010 BMC Systems Biology 4 1.

[4]   A. Capocci, V. D. P. Servedio, G. Caldarelli, and F. Colaiori, Detecting communities in large networks, 2005 Physica A 352 669.

[5]   L. Donetti and M. A. Munoz, Detecting network communities: a new systematic and efficient algorithm, 2004 Journal of Statistical Mechanics: Theory and Experiment 2004 P10012.

[6]   D. Jin, B. Yang, C. Baquero, D. Liu, D. He, and J. Liu, A Markov random walk under constraint for discovering overlapping communities in complex networks, 2011 Journal of Statistical Mechanics: Theory and Experiment 2011 P05031.

[7]   V. Zlatić, A. Gabrielli, and G. Caldarelli, Topologically biased random walk and community finding in networks, 2010 Physical Review E 82 066109.

[8]   X. Li, M. Li, Y. Hu, Z. Di, and Y. Fan, Detecting community structure from coherent oscillation of excitable systems, 2010 Physica A 389 164.

[9]   W.-J. Yuan and C. Zhou, Interplay between structure and dynamics in adaptive complex networks: Emergence and amplification of modularity by adaptive dynamics, 2011 Physical Review E 84 016116.

[10]  J. Wu, L. Jiao, C. Jin, F. Liu, M. Gong, R. Shang, and W. Chen, Overlapping community detection via network dynamics, 2012 Physical Review E 85 016115.

[11]  L. Šubelj and M. Bajec, Robust network community detection using balanced propagation, 2011 Eur. Phys. J. B 81 353.

[12]  M. J. Barber and J. W. Clark, Detecting network communities by propagating labels under constraints, 2009 Physical Review E 80 026129.

[13]  U. N. Raghavan, R. Albert, and S. Kumara, Near linear time algorithm to detect community structures in large-scale networks, 2007 Physical Review E 76 036106.

[14]  X. Liu and T. Murata, Advanced modularity-specialized label propagation algorithm for detecting communities in networks, 2010 Physica A 389 1493.

[15] B. Karrer and M. E. J. Newman, Stochastic blockmodels and community structure in networks, 2011 Physical Review E 83 016107.

[16] T. Heimo, J. M. Kumpula, K. Kaski, and J. Saramäki, Detecting modules in dense weighted networks with the Potts method, 2008 Journal of Statistical Mechanics: Theory and Experiment 2008 P08007.

[17] C. Tanmoy, Leveraging disjoint communities for detecting overlapping community structure, 2015 Journal of Statistical Mechanics: Theory and Experiment 2015 P05017.

[18] Y. Chen, X. L. Wang, B. Yuan, and B. Z. Tang, Overlapping community detection in networks with positive and negative links, 2014 Journal of Statistical Mechanics: Theory and Experiment 2014 P03021.

[19] B. Yan and S. Gregory, Detecting community structure in networks using edge prediction methods, 2012 Journal of Statistical Mechanics: Theory and Experiment 2012 P09008.

[20] J. P. Bagrow, Evaluating local community methods in networks, 2008 Journal of Statistical Mechanics: Theory and Experiment 2008 P05001.

[21] L. Danon, A. Díaz-Guilera, J. Duch, and A. Arenas, Comparing community structure identification, 2005 Journal of Statistical Mechanics: Theory and Experiment 2005 P09008.

[22] M. E. J. Newman and M. Girvan, Finding and evaluating community structure in networks, 2004 Phys. Rev. E 69 026113.

[23] A. Medus, G. Acuña, and C. O. Dorso, Detection of community structures in networks via global optimization, 2005 Physica A 358 593.

[24] G. Agarwal and D. Kempe, Modularity-maximizing graph communities via mathematical programming, 2008 Eur. Phys. J. B 66 409.

[25] V. D. Blondel, J.-L. Guillaume, R. Lambiotte, and E. Lefebvre, Fast unfolding of communities in large networks, 2008 Journal of Statistical Mechanics: Theory and Experiment 2008 P10008.

[26] S. Lehmann and L. K. Hansen, Deterministic modularity optimization, 2007 Eur. Phys. J. B 60 83.

[27] Y. Hu, J. Wu, and Z. Di, Enhance the efficiency of heuristic algorithms for maximizing the modularity Q, 2009 Europhysics Letters 85 18009.

[28] Y. Sun, B. Danila, K. Josić, and K. E. Bassler, Improved community structure detection using a modified fine-tuning strategy, 2009 Europhysics Letters 86 28004.

[29] B. H. Good, Y.-A. de Montjoye, and A. Clauset, Performance of modularity maximization in practical contexts, 2010 Physical Review E 81 046106.

[30] X. S. Zhang, R. S. Wang, Y. Wang, J. Wang, Y. Qiu, L. Wang, and L. Chen, Modularity optimization in community detection of complex networks, 2009 Europhysics Letters 87 38002.

[31] S. Fortunato and M. Barthélemy, Resolution limit in community detection, 2007 Proc. Natl. Acad. Sci. USA 104 36.

[32] J. Ruan and W. Zhang, Identifying network communities with a high resolution, 2008 Physical Review E 77 016104.

[33] D. Lai, H. Lu, and C. Nardini, Enhanced modularity-based community detection by random walk network preprocessing, 2010 Physical Review E 81 066118.

[34] J. Xiang, K. Hu, and Y. Tang, A class of improved algorithms for detecting communities in complex networks, 2008 Physica A 387 3327.

[35] J. W. Berry, B. Hendrickson, R. A. LaViolette, and C. A. Phillips, Tolerating the community detection resolution limit with edge weighting, 2011 Physical Review E 83 056119.

[36] T. Zhou, L. Lü, and Y.-C. Zhang, Predicting missing links via local information, 2009 Eur. Phys. J. B 71 623.

[37] G. Salton and M. J. McGill, Introduction to Modern Information Retrieval, McGraw-Hill, Auckland, 1983.

[38] P. Jaccard, Étude comparative de la distribution florale dans une portion des Alpes et des Jura, 1901 Bull. Soc. Vaud. Sci. Nat. 37 547.

[39] T. Sørensen, A method of establishing groups of equal amplitude in plant sociology based on similarity of species content and its application to analyses of the vegetation on Danish commons, 1948 Biol. Skr. 5 1.

[40] E. Ravasz, A. L. Somera, D. A. Mongru, Z. N. Oltvai, and A. L. Barabasi, Hierarchical organization of modularity in metabolic networks, 2002 Science 297 1551.

[41] E. A. Leicht, P. Holme, and M. E. J. Newman, Vertex similarity in networks, 2006 Physical Review E 73 026120.

[42] L. A. Adamic and E. Adar, Friends and neighbors on the Web, 2003 Social Networks 25 211.

[43] L. Lü, C.-H. Jin, and T. Zhou, Similarity index based on local paths for link prediction of complex networks, 2009 Physical Review E 80 046122.

[44] M. Girvan and M. E. J. Newman, Community structure in social and biological networks, 2002 Proc. Natl. Acad. Sci. USA 99 7821.

[45] M. E. J. Newman, Fast algorithm for detecting community structure in networks, 2004 Phys. Rev. E 69 066133.

[46] M. Rosvall and C. T. Bergstrom, Maps of random walks on complex networks reveal community structure, 2008 Proc. Natl. Acad. Sci. USA 105 1118.

[47] G. Steve, Finding overlapping communities in networks by label propagation, 2010 New Journal of Physics 12 103018.

[48] A. Lancichinetti, F. Radicchi, J. J. Ramasco, and S. Fortunato, Finding Statistically Significant Communities in Networks, 2011 Plos One 6 e18961.

[49] A. Lancichinetti, S. Fortunato, and F. Radicchi, Benchmark graphs for testing community detection algorithms, 2008 Physical Review E 78 046110.

[50] S. Muff, F. Rao, and A. Caflisch, Local modularity measure for network clusterizations, 2005 Physical Review E 72 056107.

[51] F. Radicchi, C. Castellano, F. Cecconi, V. Loreto, and D. Parisi, Defining and identifying communities in networks, 2004 Proc. Natl. Acad. Sci. USA 101 2658.

[52] P. Gleiser and L. Danon, Community Structure in Jazz, 2003 Adv. Complex Syst. 6 565.

[53] R. Guimerà, L. Danon, A. Díaz-Guilera, F. Giralt, and A. Arenas, Self-similar community structure in a network of human interactions, 2003 Physical Review E 68 065103.

[54] D. Bu, Y. Zhao, L. Cai, H. Xue, X. Zhu, H. Lu, J. Zhang, S. Sun, L. Ling, N. Zhang, G. Li, and R. Chen, Topological structure analysis of the protein–protein interaction network in budding yeast, 2003 Nucleic Acids Research 31 2443.